\documentstyle[12pt,fullpage,epsf,epsfig]{article}

\catcode`@=11 \@addtoreset{equation}{section} \catcode `@=12

\def\underrel#1\over#2{\mathrel{\mathop{\kern\z@#1}\limits_{#2}}}

\def\be{\begin{equation}}
\def\ee{\end{equation}}
\def\beq{\begin{eqnarray}}
\def\eeq{\end{eqnarray}}

\def\d{{D-brane} }
\def\dz{{D0-brane} }
\def\dzs{{D0-branes} }
\def\ds{{D6-brane} }
\def\dss{{D6-branes} }
\def\fd{{4-dimensional} }
\def\ol{{1-loop} }
\def\tl{{2-loop} }
\def\sym{{superYang-Mills} }
\def\dt{{\rm dt}}
\def\tr{{\rm Tr}}
\def\qa{{$Q^{\ell\alpha}_a$}}
\def\pa{{$P^{\ell\alpha}_a$}}
\def\pb{{$P^{m\beta}_b$}}
\def\pab{{$P^{m\beta}_a$}}
\def\d{{\cal D}}
\begin{document}
\thispagestyle{empty}

\rightline{hep-th/9902043}
\rightline{TIFR/TH/98-49}
\rightline{December 1998}
\begin{center}
\vspace{3 ex}
{\bf Probing D0-D6 Black Hole Configuration:\\
2-Loop Matrix Theory Correction to D0-Brane\\
Effective Action}\\
\vspace{8 ex}
Avinash Dhar $^*$\\
{\sl Department of Theoretical Physics}\\
{\sl Tata Institute of Fundamental Research}\\
{\sl Homi Bhabha Road, Mumbai 400 005, India.}\\
\vspace{15 ex}
\pretolerance=1000000
\end{center}
\begin{abstract}

We present a calculation of the matrix theory \tl effective action for
a \dz in the background of the recently discussed D0-D6 bound state
configuration. The effective DBI action of a \dz probe in the
background of the corresponding \fd non-supersymmetric black hole
solution to low-energy type IIA string theory compactified on a
6-torus is known to agree with the matrix theory calculation at \ol
order, in the limit in which the ratio of the \dz to the \ds charges
carried by the black hole is large. Agreement at \tl between the
supergravity description and a conjectured nonabelian BDI effective
\sym description has also been recently reported. However, we find
uncanceled ultraviolet divergences in our direct matrix theory
calculation of the \tl effective action. This is consistent with the
expected nondecoupling of massive open string states from the 6-brane.

\end{abstract}

\vfill
\hrule
\vspace{0.5 ex}
\leftline{$^*$ adhar@theory.tifr.res.in}
\clearpage

\section{Introduction and Summary}

Recently several authors have discussed a non-supersymmetric
configuration of D0- and \dss \cite{SHEIN, WATI, KVK, JMP, BISY, DM,
NI} in the context of supergravity/\sym correspondence conjectured by
matrix theory \cite{BFSS}. These studies have uncovered many interesting 
features of this brane configuration. This configuration appears as a 
classical solution in (6+1)-dimensional \sym theory \cite{WATI}. 
The energy spectrum of the configuration with arbitrary number of D0-
and \dss \cite{DM}, which is labelled by the corresponding
Ramond-Ramond charges, turns out to be identical to the mass spectrum of a
non-supersymmetric extremal black hole solution to the classical
equations of low-energy type IIA string theory compactified on a
six-torus down to 4-dimensions \cite{GK, LP, DM, DoM, CD, DP, GW}. 
This is surprising because in the
absence of supersymmetry in this system one would have expected the
weak coupling gauge theory result for the energy spectrum to get
modified in the strong coupling regime corresponding to
supergravity. Furthermore, the configuration has been studied with
various brane probes and agreement has been found at the \ol level 
between matrix theory and supergravity calculations \cite{KVK, JMP, 
BISY, DM, GL}, in the limit of large zero to six brane charge ratio. In
this limit, a \dz scattering off this configuration is dominated by
D0-D0 scattering and this is what explains the agreement between the
two calculations of the $v^4$ terms in the \ol effective action, even
in the absence of supersymmetry in this system. However, the agreement
of the $v^2$ and constant ($v$-independent) terms in the effective
action is difficult to understand since these terms manifestly
represent the supersymmetry breaking interactions between D0- and \ds.
Recently, a \tl calculation for a \ds probe using the effective \sym
action of Chepelev and Tsyetlin \cite{CT} has also been reported
\cite{JB} to be in agreement with the corresponding supergravity
calculation at \tl. These results are surprising not only because the
configuration is non-supersymmetric but also because one does not
expect the bulk to decouple from the 6-brane. It is, therefore, of
interest to ask whether the agreement at \tl level persists in a
direct matrix theory calculation.

In this paper we present a direct calculation in the matrix theory
framework of the \tl effective action of a \dz probe in
the background of the D0-D6 bound state configuration. We find
uncanceled divergences in the matrix theory calculation. These
divergences seem to be related to the ultraviolet divergences one
might expect in a $(6+1)$-dimensional field theory. This result, which
is expected \cite{SEN, SEI}, makes the agreement found in \cite{JB} 
all the more intriguing.

The present work draws heavily on \cite{DM} and should be regarded as a
sequel to that work. We will follow the conventions and notations of
this reference. We also refer the reader to this work for all the
background material which we will need here. The organization of this
paper is as follows. In the next section we first briefly discuss the
D0-D6 configuration given in \cite{DM} and then give the terms in the matrix
theory action, expanded around this configuration, which are relevant
to the \tl calculation. In section 3 we give the various 2-point
functions needed for the calculation. The calculation of the \tl
effective action is presented in section 4. In section 5 we discuss
the origin of the uncanceled divergences in the \tl effective
action. We end with some concluding
remarks in section 6. Expressions for the various terms in the \tl
effective action are collected together in the Appendix.

\section{The Matrix Theory Action}

In this section we will expand the matrix theory action around the
D0-D6 background configuration discussed in section 5 of \cite{DM}, and
briefly reviewed below, and obtain the terms relevant for a \tl
calculation. The matrix theory action is
\beq 
S &=&\frac{1}{2g} \int \dt~\tr \bigg\{ (D_tX_i)^2 + \frac{1}{2}
[X_i,X_j]^2 - (\bar D_tA)^2 + \theta^T D_t \theta \nonumber \\ [2mm] 
&&+ i \theta^T \gamma^i [X_i, \theta] + 2 \partial_t C^\dagger D_t C 
- 2 [C^\dagger,B_i] [X_i,C] \bigg\} 
\eeq 
The notations and conventions are the same as in \cite{DM}.

\bigskip

\noindent\underbar{\bf Background Configuration}

The D0-D6 background configuration consists of multiple \dss with the
\dzs appearing on them as magnetic fluxes. We present this
configuration below:
\[
B_{4,6,8}=\left[\matrix{&&\vdots& \cr Q_{1,2,3}&&\vdots&0\cr
\cdots&\cdots&\cdots&\cr 0&&\vdots&0}\right] , \ \ 
B_{5,7,9}=\left[\matrix{&&\vdots& \cr P_{1,2,3}&&\vdots&0\cr 
\cdots&\cdots&\cdots&\cr 0&&\vdots&0}\right]
\]
Here, the entry in the lower right corner is a single-element
one. This entry is reserved for the probe \dz. For a single D6-brane,
$[Q_a, P_a] = ic_a$, $a = 1,2,3$. For multiple parallel D6-branes, the
$Q_a$'s and the $P_a$'s have a further structure:
\[
Q_a = \left[ \matrix{Q^{(1)}_a &&& \cr & Q^{(2)}_a && \cr && \ddots &
\cr &&& Q^{(Q_6)}_a}\right] , \ \ 
P_a = \left[ \matrix{P^{(1)}_a &&& \cr & P^{(2)}_a && \cr && \ddots &
\cr &&& P^{(Q_6)}_a}\right] , 
\]
where $[Q^{(1)}_a , P^{(1)}_a] = ic^{(1)}_a$, etc. The \dss are wrapped
on $T^6$ with volume $V_6$ which is assumed to be large since we will
be neglecting the effect of winding modes. 

The $Q_6$ \dss are organized into four sets, each consisting of $n$
\dss, where $4n=Q_6$. Each \ds in the first set carries magnetic
fluxes $(F_{45} , F_{67} , F_{89}) = (f , f , f)$. The \dss in the
other three sets carry the magnetic fluxes $(f , -f , -f)$, $(-f , -f
, f)$ and $(-f , f , -f)$. Here $f$ is the parameter which is related
to the ratio of the \dz to \ds charge in the bound state \cite{DM}. In
view of the above, a more suitable notation for the $Q_a$'s and
$P_a$'s is \qa, \pa, where $\ell = 1,2,3,4$ and $\alpha = 1, 2,
\cdots, n$ and [\qa, \pb] $= i\delta_{ab} \delta^{\alpha\beta}
\delta^{\ell m} c^a_{\ell\alpha}$. For the desired configuration we
need $c^a_{\ell\alpha} = c^a_\ell$ to be independent of
$\alpha$. Moreover, the four triplets of numbers $\{c^a_\ell\} =
(c^1_\ell , c^2_\ell , c^3_\ell) \equiv \vec c_\ell$ correspond to the
four triplets of fluxes listed above and so we may write
\be
\vec c_\ell = c \vec\epsilon_\ell,~~\vec\epsilon_1 = (1,1,1),
~~\vec\epsilon_2 = (1,-1,-1),~~\vec\epsilon_3 = (-1,-1,1),
~~\vec\epsilon_4 = (-1,1,-1).
\ee
Agreement with supergravity at \ol requires $c = f^{-1} \rightarrow
0$. 

A \dz scattering off this background in a direction transverse to the
\dss is represented by the additional background fields
\[
B_1=\left[\matrix{&&\vdots& \cr 0&&\vdots&0\cr
\cdots&\cdots&\cdots&\cr 0&&\vdots&vt}\right] , \ \ 
B_2=\left[\matrix{&&\vdots& \cr 0&&\vdots&0\cr 
\cdots&\cdots&\cdots&\cr 0&&\vdots&b}\right] , \ \ B_3 = 0 .
\]
The only nonzero entry in the above matrices is in the lower right
corner. Here $v$ is the velocity of the D0-brane, assumed to be moving
along the $x^1$ direction, and $b$ is the impact parameter.

\bigskip

\noindent\underbar{\bf Fluctuations}

There are basically two different types of fluctuations that we need
to consider. The first type are the ones with nonzero entries only in
the last row or column of the various matrix variables. These
represent open strings connecting the probe \dz to the branes in the
background. These type of fluctuations are the only ones that
contribute at the \ol level. We shall call them ``column type''
fluctuations. The other type of fluctuations have nonzero entries
everywhere other than the last row and the last column. These ``matrix
type'' fluctuations represent open strings connecting the various
branes in the background. They do not contribute to the \ol effective
action, but do contribute at \tl level and beyond. We have done the
\tl computation in the limit of a large number of \ds in the
background configuration, i.e. for large values of $Q_6$. The
fluctuation in the bottom right corner entry in the various matrix
variables can be ignored for the leading contribution in this
limit. This is what we have done in the following.

Writing $X_i = B_i + \sqrt g Y_i$, we have
\[
Y_i=\left[\matrix{&&\vdots& \cr Z_i&&\vdots&\phi_i\cr
\cdots&\cdots&\cdots&\cr \phi^\dagger_i&&\vdots&0}\right] 
\]
Similarly,
\[
A = \sqrt g\left[\matrix{&&\vdots& \cr Z_A&&\vdots&\phi_A\cr
\cdots&\cdots&\cdots&\cr \phi^\dagger_A&&\vdots&0}\right] , \ \ 
\Theta = \sqrt g\left[\matrix{&&\vdots& \cr Z_\theta&&\vdots&\chi_\theta\cr
\cdots&\cdots&\cdots&\cr \chi^\dagger_\theta&&\vdots&0}\right] , \ \ 
C = \sqrt g\left[\matrix{&&\vdots& \cr Z_c&&\vdots&\phi_c\cr
\cdots&\cdots&\cdots&\cr \tilde\phi^T_c&&\vdots&0}\right] 
\]
Note that because $C$ is not hermition, $\tilde\chi_c \neq
\chi^\ast_c$. Also, in the notation that we have used for the
background, the fluctuations have the index structure
$\phi^{\ell\alpha}_{\cdots} , \chi^{\ell\alpha}_{\cdots} ,
Z^{\ell\alpha, m\beta}_{\cdots}$. Finally, it is useful to parameterize
the background as 
\[
B_i=\left[\matrix{&&\vdots& \cr D_i&&\vdots&0\cr
\cdots&\cdots&\cdots&\cr 0&&\vdots&d_i}\right] . 
\]

\bigskip

\noindent\underbar{\bf Action for Fluctuations}

The action (2.1) may now be expanded around the background $B_i$. It
is convenient to write the result as follows:
\be
S = S_{Bgd} + S_2 + S_{\rm Int} .
\ee
Here $S_2$ is the part of the action quadratic in fluctuations and
$S_{\rm Int}$ is the part containing interactions. Furthermore, it is
convenient to write
\be
S_2 = S^{X+A}_2 + S^\theta_2 + S^C_2 .
\ee
We find,
\beq
S^{X+A}_2 &=& \int d\tau \ \bigg\{{1\over 2} \tr \left( \dot Z^2_i -
[\d_i, Z_j]^2 - 2 [\d_i, \d_j] [Z_i, Z_j]\right) \nonumber \\ [2mm]
&& ~ + \dot\phi^\dagger_i \dot\phi_i + \phi^\dagger_j \d^2_i \phi_j +
2 \phi^\dagger_i [\d_i, \d_j] \phi_j \nonumber \\ [2mm]
&& ~ + {1\over 2} \tr \left( \dot Z^2_A - [\d_i, Z_A]^2\right) +
\dot\phi^\dagger_A \dot\phi_A \nonumber \\ [2mm]
&& ~ + \phi^\dagger_A \d^2_i \phi_A + 2i \dot d_i \left(\phi^\dagger_i
\phi_A - \phi^\dagger_A \phi_i\right) \bigg\}
\eeq
\be
S^\theta_2 = \int d\tau \ \bigg\{{i\over 2} \tr \left( Z^T_\theta \dot
Z_\theta - Z^T_\theta \gamma^i [\d_i, Z_\theta]\right) + i
\chi^\dagger_\theta \dot\chi_\theta - i \chi^\dagger_\theta \gamma^i \d_i
\chi_\theta \bigg\} ,
\ee
\be
S^C_2 = \int d\tau \ \bigg\{ \tr \left( \dot Z^\dagger_c \dot Z_c -
[\d_i , Z^\dagger_c] [\d_i Z_c]\right) + \dot\phi^\dagger_c \dot\phi_c
+ \phi^\dagger_c \d_i^2 \phi_c + \dot{\tilde \phi}^\dagger_c
\dot{\tilde\phi}_c + \tilde\phi^\dagger_c \d_i^2 \tilde\phi_c \} .
\ee
In writing the above, we have already made a Wick rotation to
Euclidean time, $t \rightarrow i\tau$, $A \rightarrow -iA$ and $vt
\rightarrow v_E\tau$.\footnote{Here $v_E = iv$ is the Euclidean
velocity, to be taken real during the course of this calculation. For
convenience of notation, we will drop the subscript $`E$' in the
following.} Also, a dot represents derivative with respect $\tau$, and
we have used the notation
\be
\d_i \equiv D_i - d_i {\bf 1}
\ee

Similarly to the quadratic piece, it is convenient to write the
interaction piece of the action for fluctuations as 
\be
S_{\rm Int} = S^{X+A}_{\rm Int} + S^\theta_{\rm Int} + S^C_{\rm
Int} .
\ee
We find
\beq
S^{X+A}_{\rm Int} &=& \sqrt{g} \int d\tau \
\bigg\{\left(\phi^\dagger_i \d_i Z_j \phi_j + \phi^\dagger_j \d_i Z_i
\phi_j - 2\phi^\dagger_j \d_i Z_j \phi_i + h.c.\right) \nonumber \\ [2mm]
&&\hspace{1cm} + ~ \left(\phi^\dagger_i \d_i Z_A \phi_A +
\phi^\dagger_A \d_i Z_i 
\phi_A - 2\phi^\dagger_A \d_i Z_A \phi_i + h.c.\right) \nonumber \\ [2mm]
&&\hspace{1cm} + ~ \left( i\phi^\dagger_i Z_i \dot\phi_A + i \phi^\dagger_i Z_A
\dot\phi_i + 2i \dot\phi^\dagger_i Z_i \phi_A + h.c. \right)\bigg\}
\nonumber \\ [2mm]
&& +g \int d\tau \ \bigg\{ \phi^\dagger_i Z_i Z_j \phi_j + \phi^\dagger_j
Z^2_i \phi_j - 2 \phi^\dagger_i Z_j Z_i \phi_j \nonumber \\ [2mm]
&&\hspace{1cm} - ~ (\phi^\dagger_i \phi_j)^2 + {1\over 2}
(\phi^\dagger_i \phi_i)^2 + {1\over 2} (\phi^\dagger_i \phi_j)
(\phi^\dagger_j \phi_i) \nonumber \\ [2mm]
&&\hspace{1cm} + ~ \phi^\dagger_i Z^2_A \phi_i + \phi^\dagger_A
Z^2_i \phi_A - (\phi^\dagger_A \phi_i)^2 - (\phi^\dagger_i \phi_A)^2
\nonumber \\ [2mm] 
&&\hspace{1cm} + ~ (\phi^\dagger_A \phi_A) (\phi^\dagger_i \phi_i) +
(\phi^\dagger_A \phi_i) (\phi^\dagger_i \phi_A)\bigg\} ,
\eeq
\be
S^\theta_{\rm Int} = \sqrt{g} \int d\tau \ \bigg\{ -i
\chi^\dagger_\theta \gamma^i Z_i \chi_\theta + i \chi^\dagger_\theta \gamma^i
Z_\theta \phi_i + i \phi^\dagger_i Z_\theta \gamma^i \chi_\theta - 
\chi^\dagger_\theta Z_\theta \phi_A - \phi^\dagger_A
Z_\theta \chi_\theta + \chi^\dagger_\theta Z_A \chi_\theta \bigg\} ,
\ee
\beq
S^C_{\rm Int} &=& \sqrt{g} \int d\tau \ \bigg\{ \tilde\phi^T_c [\d_i,
Z^\dagger_c] \phi_i + \phi^\dagger_i [\d_i, Z^\dagger_c] \phi_c -
\tilde\phi^T_c Z_i \d_i \tilde\phi^\ast_c +
\phi^\dagger_i Z_c \d_i \tilde\phi^\ast_c \nonumber \\ [2mm]
&&\hspace{1cm} + ~ \phi^\dagger_c \d_i Z_i \phi_c - \phi^\dagger_c
\d_i Z_c \phi_i - i \bigg( \dot{\tilde\phi}^T_c Z^\dagger_c
\phi_A + \tilde\phi^T_c Z_c \dot\phi_A + \dot\phi^\dagger_A
Z^\dagger_c \phi_c \nonumber \\ [2mm]
&&\hspace{1cm} + ~ \phi^\dagger_A Z^\dagger_c \dot\phi_c +
\tilde\phi^T_c Z_A \dot{\tilde\phi}^\ast_c + \dot\phi^\dagger_c Z_A
\phi_c  - \phi^\dagger_A Z_c \dot{\tilde\phi}^\ast_c -
\dot\phi^\dagger_c Z_c \phi_A \bigg) \bigg\} .
\eeq
In the above, `h.c.' stands, as usual, for hermitian conjugate. Also, we
have dropped all terms that either do not contribute to the \tl
effective action of the probe \dz in the leading large $Q_6$ limit or
contribute terms that do not depend on the impact parameter $b$ and
the velocity $v$ of the probe.

Note that in the notation we have used for the background, a term
like, for example, $\tr \dot Z^2_i$ reads $\sum^4_{\ell, m = 1} \
\sum^n_{\alpha, \beta = 1} tr (\dot Z^{\ell \alpha , m \beta}_i \ \dot
Z^{m \beta , \ell\alpha}_i)$. The remaining trace $`tr$' is over the
space in which the harmonic oscillator variables \qa , \pa operate.

\section{Two-Point Functions}

There are two different kinds of 2-point functions corresponding to
the two different kinds of fluctuations, namely ``column type''
fluctuations $\phi_i , \phi_A , \chi_\theta , \phi_c$ and
$\tilde\phi_c$ and ``matrix type'' fluctuations $Z_i , Z_A , Z_\theta$
and $Z_c$. One minor complication in both the sectors is that there is
mixing at the quadratic level, so the propagators cannot be
immediately read-off from the quadratic action, $S_2$, and a  
rediagonalization is needed. Furthermore, to obtain explicit
expressions for the 2-point functions, it is necessary to use an
explicit representation for the Heisenberg algebra, [\qa , \pab ] = $ic
\delta^{\ell m} \delta^{\alpha\beta} \varepsilon^a_\ell$. In the limit
$c \rightarrow 0$, we may use the representation in terms of functions
of 3 real variables, one for each of the three values of the index
$`a$'. We will denote these three real variables by the triplet $\vec
x \equiv (x_1, x_2, x_3)$. Note that in terms of there variables the
explicit form for a term like, for example, \tr $\dot Z^2_i$ is
$\sum^4_{\ell, m=1} \ \sum^n_{\alpha,\beta=1} \int d^3 \vec x \int d^3
\vec y \ |\dot Z^{\ell\alpha, m\beta}_i (\vec x, \vec y, \tau)|^2$. We
will now list below all the nonzero 2-point functions. It is
convenient to use the proper time representation and this is what we
do below.

\bigskip

\noindent\underbar{\bf Column Type Fluctuations}

Let us first consider the ``column type'' fluctuations, $\phi_i ,
\phi_A$, etc. Introducing the notation
\be
\langle \phi^{\ell\alpha}_I (\vec x, \tau) \ \phi^{m\beta\ast}_J
(\vec x', \tau') \rangle \equiv \delta^{\ell m} \delta^{\alpha\beta}
\Delta^{\ell\alpha}_{IJ} (\vec x, \tau ; \vec x', \tau') ,
\ee
where $I, J \epsilon (i, A, c, \tilde c, \theta)$\footnote{Here the 
subscript $\tilde c$ refers to $\tilde\phi_c$.}, we have
\beq
\Delta^{\ell\alpha}_{AA} = \Delta^{\ell\alpha}_{11} &=& \int^\infty_0
ds \ \cosh 2vs \ K_s (\vec x, \tau ; \vec x' \tau') , \\ [2mm]
\Delta^{\ell\alpha}_{A1} = - \Delta^{\ell\alpha}_{1A} &=& i \int^\infty_0
ds \ \sinh 2vs \ K_s (\vec x, \tau ; \vec x' \tau') , \\ [2mm]
\Delta^{\ell\alpha}_{22} = \Delta^{\ell\alpha}_{33} =
\Delta^{\ell\alpha}_{cc} = \Delta^{\ell\alpha}_{\tilde c \tilde c} &=&
\int^\infty_0 ds \ K_s (\vec x, \tau ; \vec x' \tau') , \\ [2mm]
\Delta^{\ell\alpha}_{ii} &=& \int^\infty_0 ds \ \cosh 2cs \ K_s (\vec x,
\tau ; \vec x', \tau'), \ (i = 4, 5, \cdots , 9) \\ [2mm]
\Delta^{\ell\alpha}_{45} = - \Delta^{\ell\alpha}_{54} &=& - i
\epsilon^1_\ell \int^\infty_0 ds \ \sinh 2cs \ K_s (\vec x, \tau ; \vec
x', \tau') , \\ [2mm]
\Delta^{\ell\alpha}_{67} = - \Delta^{\ell\alpha}_{76} &=& - i
\epsilon^2_\ell \int^\infty_0 ds \ \sinh 2cs \ K_s (\vec x, \tau ; \vec
x', \tau') , \\ [2mm]
\Delta^{\ell\alpha}_{89} = - \Delta^{\ell\alpha}_{98} &=& - i
\epsilon^3_\ell \int^\infty_0 ds \ \sinh 2cs \ K_s (\vec x, \tau ; \vec
x', \tau') , \\ [2mm]
\Delta^{\ell\alpha}_{\theta\theta} &=& i \left(\partial_\tau + \gamma^j
\d^{\ell\alpha}_{jx}\right) \int^\infty_0 ds \ \exp \left[-s (v\gamma^1 +
c\vec\sigma \cdot \vec\epsilon_\ell)\right] \ K_s (\vec x, \tau ; \vec
x', \tau'). \nonumber \\ [2mm]. 
\eeq
In the above
\beq
K_s (\vec x, \tau ; \vec x', \tau') &=& {1 \over 4\pi^2} (v/c^3 \sinh^32cs
\sinh 2vs)^{1/2} \ e^{-sb^2} \times \nonumber \\ [2mm]
&& exp \bigg[- {v \over 4} \left\{(\tau + \tau')^2 \tanh vs 
+ (\tau - \tau')^2 \coth vs  \right\} \nonumber \\ [2mm]
&& \hspace{1cm} - {1\over 4c} \left\{ (\vec x + \vec x')^2 \tanh cs +
(\vec x - \vec x')^2 \coth cs \right\} \bigg] .
\eeq
Also, we have used the following notation in (3.9):
\beq
&& \d^{\ell\alpha}_{1x} = -v\tau , \ \ \d^{\ell\alpha}_{2x} = -b , \ \
\d^{\ell\alpha}_{3x} = 0 , \nonumber \\ [2mm]
&& \left(\d^{\ell\alpha}_{4x} , \d^{\ell\alpha}_{6x} ,
\d^{\ell\alpha}_{8x}\right) = (x_1, x_2, x_3) , \nonumber \\ [2mm]
&& \left(\d^{\ell\alpha}_{5x} , \d^{\ell\alpha}_{7x} ,
\d^{\ell\alpha}_{9x}\right) = -ic \left(\epsilon^1_\ell {\partial
\over \partial x_1} , \epsilon^2_\ell {\partial \over \partial x_2}
, \epsilon^3_\ell {\partial \over \partial x_3}\right) ,
\eeq
and
\be
\vec\sigma = (\sigma^1 , \sigma^2 , \sigma^3) \equiv (i \gamma^4 \gamma^5 , i
\gamma^6 \gamma^7 , i \gamma^8 \gamma^9) .
\ee

\newpage

\noindent\underbar{\bf Matrix Type Fluctuations}

We now consider the matrix type fluctuations, $Z_i, Z_A$, etc. We
introduce a notation similar to (3.1),
\be
\langle Z^{\ell\alpha, m\beta}_I (\vec x, \vec y, \tau) \
Z^{\ell'\alpha', m'\beta'\ast}_J
(\vec x', \vec y', \tau') \rangle \equiv \delta^{\ell\ell'}
\delta^{mm'} \delta^{\alpha\alpha'} \delta^{\beta\beta'} 
\Lambda^{\ell\alpha , m\beta}_{IJ} (\vec x, \vec y, \tau ; \vec x', \vec
y', \tau') ,
\ee
where $I, J \epsilon (i, A, c, \theta)$. A complication here is
that the cases (i) $\ell = m$ and (ii) $\ell \neq m$ need to be
considered separately since the 2-point functions are different in the
two cases. This is because there is mixing at the quadratic level in
case (ii). The 2-point functions, however, depend only on the
combination $\{ \epsilon^a_{\ell m} \} \equiv \{ \epsilon^a_\ell
\epsilon^a_m\}$ and therefore only four independent cases need to
be considered, instead of a possible sixteen. We now list all the
nonvanishing 2-point functions in their proper time representation. 

\noindent(i)~ \underbar{$\ell = m$} 

\be
\Lambda^{\ell\alpha, \ell\beta}_{\theta\theta} = i \left\{
\partial_\tau + \gamma^j \left( \d^{\ell\alpha}_{jx} -
\d^{\ell\beta\ast}_{jy}\right)\right\} \tilde\Lambda^{\ell\alpha,
\ell\beta}_{\theta\theta} ,
\ee
\beq
\Lambda^{\ell\alpha, \ell\beta}_{ii} &=& \Lambda^{\ell\alpha,
\ell\beta}_{AA} = \Lambda^{\ell\alpha, \ell\beta}_{cc} =
\tilde\Lambda^{\ell\alpha, \ell\beta}_{\theta\theta} \nonumber \\ [2mm]
&=& {1\over 16\pi^2 c^3} \int^\infty_0 {ds \over s^2} \exp \left[ -s
(\vec x - \vec y)^2 - {(\vec x - \vec x')^2 \over 4c^2 s} -
{(\tau - \tau')^2 \over 4s}\right] \delta^{(3)} \left((\vec x -
\vec x') - (\vec y - \vec y')\right) . \nonumber \\
\eeq

\bigskip

\noindent(ii)~ \underbar{$\ell \neq m$}

\beq
\Lambda^{\ell\alpha, m\beta}_{ii} (i = 1,2,3) &=& \Lambda^{\ell\alpha,
m\beta}_{AA} = \Lambda^{\ell\alpha, m\beta}_{cc} = \int^\infty_0 ds \
G^{\{\epsilon^a_{\ell m}\}}_s (\vec x, \vec y, \tau; \vec x', \vec
y', \tau') , \\ [2mm]
\Lambda^{\ell\alpha, m\beta}_{44} = \Lambda^{\ell\alpha,
m\beta}_{55} &=& \int^\infty_0 ds \ \cosh \left[2cs(1-\epsilon^1_{\ell
m})\right] G^{\{\epsilon^a_{\ell m}\}}_s (\vec x, \vec y, \tau;
\vec x', \vec y', \tau') , \\ [2mm]
\Lambda^{\ell\alpha, m\beta}_{45} = - \Lambda^{\ell\alpha,
m\beta}_{54} &=& -i \epsilon^1_\ell \int^\infty_0 ds \ \sinh
\left[2cs(1-\epsilon^1_{\ell m})\right] G^{\{\epsilon^a_{\ell
m}\}}_s (\vec x, \vec y, \tau; \vec x', \vec y', \tau') , \\ [2mm]
\Lambda^{\ell\alpha, m\beta}_{66} = \Lambda^{\ell\alpha,
m\beta}_{77} &=& \int^\infty_0 ds \ \cosh \left[2cs(1-\epsilon^2_{\ell
m})\right] G^{\{\epsilon^a_{\ell m}\}}_s (\vec x, \vec y, \tau;
\vec x', \vec y', \tau') , \\ [2mm] 
\Lambda^{\ell\alpha, m\beta}_{67} = \Lambda^{\ell\alpha,
m\beta}_{76} &=& -i \epsilon^2_\ell \int^\infty_0 ds \ \sinh
\left[2cs(1-\epsilon^2_{\ell m})\right] G^{\{\epsilon^a_{\ell
m}\}}_s (\vec x, \vec y, \tau; \vec x', \vec y', \tau') , \\ [2mm]
\Lambda^{\ell\alpha, m\beta}_{88} = \Lambda^{\ell\alpha,
m\beta}_{99} &=& \int^\infty_0 ds \ \cosh \left[2cs(1-\epsilon^3_{\ell
m})\right] G^{\{\epsilon^a_{\ell m}\}}_s (\vec x, \vec y, \tau;
\vec x', \vec y', \tau') , \\ [2mm]
\Lambda^{\ell\alpha, m\beta}_{89} = \Lambda^{\ell\alpha,
m\beta}_{98} &=& -i \epsilon^3_\ell \int^\infty_0 ds \ \sinh
\left[2cs(1-\epsilon^3_{\ell m})\right] G^{\{\epsilon^a_{\ell
m}\}}_s (\vec x, \vec y, \tau; \vec x', \vec y', \tau') , 
\eeq
\be
\Lambda^{\ell\alpha, m\beta}_{\theta\theta} = i \left\{\partial_\tau +
\gamma^j (\d^{\ell\alpha}_{jx} - \d^{m\beta\ast}_{jy})\right\}
\int^\infty_0 ds \ \exp \left[-cs (\vec\epsilon_\ell -
\vec\epsilon_m)\cdot \vec\sigma\right] G^{\{\epsilon^a_{\ell
m}\}} (\vec x, \vec y, \tau; \vec x', \vec y', \tau').
\ee
Note that for the choice of $\vec\epsilon_\ell$'s given in (2.2),
$\{\epsilon^a_{\ell m}\} , \ell \neq m$, always has one +ve and two
-ve signs. For $\{\epsilon^a_{\ell m}\} = (+, -, -)$ we have

\newpage

\beq
G^{(+,-,-)}_s &=& (4\pi^2 c^2s \ \sinh 4cs)^{-1} \exp \left[-s (x_1-y_1)^2
- {(x_1-x_1')^2 \over 4c^2s} - {(\tau - \tau')^2 \over 4s}\right]
\nonumber \\ [2mm]
&&\times \exp \bigg[- {1\over 8c}  \left\{\left((x_2 + x_2') -
(y_2+y_2')\right)^2 + \left((x_3+x_3') - (y_3+y_3')\right)^2
\right\} \tanh2cs \nonumber \\ [2mm]
&&\hspace{1.5cm}  - {1\over 8c} \left\{\left((x_2-x_2') -
(y_2-y_2')\right)^2 + \left((x_3-x_3') - (y_3-y_3')\right)^2
\right\} \coth2cs\bigg] \nonumber \\[2mm] 
&&\times \delta \left((x_1-x_1') - (y_1-y_1')\right) \delta
\left((x_2-x_2') + (y_2-y_2')\right) \delta \left((x_3 - x_3') +
(y_3-y_3')\right) . \nonumber \\
\eeq
$G^{(-, +, -)}_s$ is obtained from the above by the substitutions $x_1
\leftrightarrow x_2$, $y_1 \leftrightarrow y_2$ and $x_1'
\leftrightarrow x_2'$, $y_1' \leftrightarrow y_2'$. Similarly,
$G^{(-, -, +)}_s$ is obtained by the substitutions $x_1
\leftrightarrow x_3$, $y_1 \leftrightarrow y_3$ and $x_1'
\leftrightarrow x_3'$ $y_1' \leftrightarrow y_3'$.

Using the 2-point functions listed above and the triple and quartic
interaction vertices given in (2.10) -- (2.12), standard rules of
perturbation theory may be used to evaluate the \tl diagrams.

\section{The Two-Loop Effective Action}

There are basically two types of 1PI diagrams that contribute to the
effective action at the \tl level. Diagrams of the type shown in
Fig. 1 involve two 3-point vertices and diagrams of the type shown
\begin{figure}[htb]
\begin{center}
\leavevmode
\hbox{%
\epsfxsize=5.8in
\epsffile{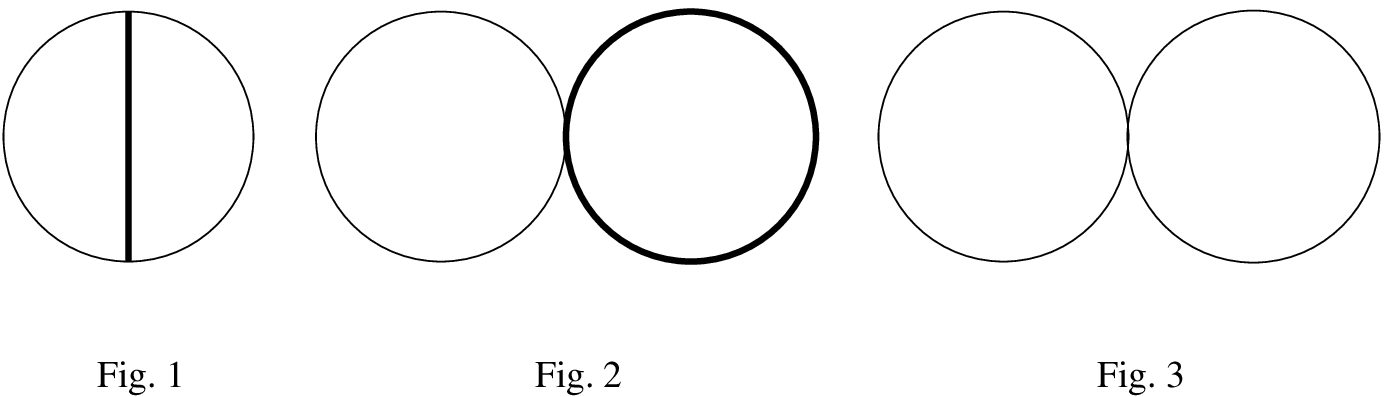}}
\end{center}
\end{figure}
in Figs. 2 and 3 involve a single 4-point vertex. In these diagrams
the thin lines represent ``column type'' propagators and the thick 
lines represent ``matrix type'' propagators. Diagrams of
type Figs. 2 and 3 are easier to evaluate and so we compute these
first. 

\bigskip

\noindent\underbar{\bf Diagrams Involving a 4-Point Vertex}

\begin{enumerate}
\item[{(i)}] Diagrams involving a ``matrix type'' propagator, Fig. 2,
come from terms of order $g$ in $S^{X+A}_{\rm Int}$, (2.10),
containing two $Z$'s. There are five such terms. Their contribution is
\beq
\Gamma_Z &=& g \sum^4_{\ell, m=1} \sum^n_{\alpha, \beta = 1} \int d\tau
\ \int d^3\vec x \ \int d^3 \vec y \ \int d^3 \vec z \ \bigg[ -2
\Delta^{\ell\alpha}_{ij} \Lambda^{\ell\alpha, m\beta}_{ij} +
\Delta^{\ell\alpha}_{ij} \Lambda^{\ell\alpha, m\beta}_{ji} \nonumber \\ [2mm]
&& + \Delta^{\ell\alpha}_{ii} \Lambda^{\ell\alpha, m\beta}_{jj} +
\Delta^{\ell\alpha}_{ii} \Lambda^{\ell\alpha, m\beta}_{AA} +
\Delta^{\ell\alpha}_{AA} \Lambda^{\ell\alpha, m\beta}_{ii} \bigg] ,
\eeq
where $\Delta^{\ell\alpha}_{IJ} \equiv \Delta^{\ell\alpha}_{IJ} (\vec
z, \tau; \vec x, \tau)$ and $\Lambda^{\ell\alpha, m\beta}_{IJ} \equiv
\Lambda^{\ell\alpha, m\beta}_{IJ} (\vec x, \vec y, \tau ; \vec z, \vec
y, \tau)$. We have evaluated this using the proper time
representations for the 2-point functions involved. The result is
given in the Appendix in (A6).

\item[{(ii)}] Diagrams involving only ``column type'' propagators,
Fig. 3, come from the rest of the order $g$ terms in (2.10). There are
seven such terms. Their contribution is 
\beq
\Gamma_\phi &=& g \sum^4_{\ell, m=1} \sum^n_{\alpha, \beta = 1} \int d\tau
\ \int d^3\vec x \ \int d^3 \vec y \ \bigg[ -\Delta^{\ell\alpha}_{ij}
\Delta^{m\beta}_{ij} + {1\over 2} \Delta^{\ell\alpha}_{ii}
\Delta^{m\beta}_{jj}  \nonumber \\ [2mm]
&& + {1\over 2} \Delta^{\ell\alpha}_{ij} \Delta^{m\beta}_{ji} - 3 
\Delta^{\ell\alpha}_{1A} \Delta^{m\beta}_{1A} +
\Delta^{\ell\alpha}_{AA} \Delta^{m\beta}_{ii} \bigg] ,
\eeq
where $\Delta^{\ell\alpha}_{IJ} \equiv \Delta^{\ell\alpha}_{IJ} (\vec
x, \tau ; \vec x, \tau)$ and $\Delta^{m\beta}_{IJ} \equiv
\Delta^{m\beta}_{IJ} (\vec y, \tau ; \vec y, \tau)$. This has also
been evaluated and is given in the Appendix in (A7).
\end{enumerate}

\bigskip

\noindent\underbar{\bf Diagrams Involving Two 3-Point Vertices}

It is convenient to group these diagrams, Fig. 1, into sets such that
each set has only one type of matrix propagator joining the
two 3-point vertices. Thus, we have diagrams with only matrix ghost
propagator, etc. We now give below the contributions of the different
sets. Note that to arrive at the expressions listed below we have
made extensive use of the symmetry properties of the various 2-point
functions with respect to their arguments as well as their
indices. Also, throughout the following $\Delta^{\ell\alpha}_{IJ}
\equiv \Delta^{\ell\alpha}_{IJ} (\vec x', \tau' ; \vec x, \tau) ,
\Delta^{m\beta}_{IJ} \equiv \Delta^{m\beta}_{IJ} (\vec y, \tau ; \vec
y', \tau')$ and $\Lambda^{\ell\alpha, m\beta}_{IJ} \equiv
\Lambda^{\ell\alpha, m\beta}_{IJ} (\vec x, \vec y, \tau ; \vec x', \vec
y', \tau')$. 

\begin{enumerate}
\item[{(i)}] Diagrams with a matrix ghost propagator give the
following contribution to the effective action.
\beq
\Gamma_C &=& - g \sum^4_{\ell, m=1} \sum^n_{\alpha, \beta = 1} \int d\tau
\int d^3\vec x \int d^3 \vec y \int d\tau' \int d^3 \vec x' \int d^3
\vec y' \nonumber \\ [2mm] 
&&\hspace{1cm} \bigg[ 2 \left( \partial_\tau \Delta^{\ell\alpha}_{cc}
\Delta^{m\beta}_{AA} - iv\tau \Delta^{\ell\alpha}_{cc}
\Delta^{m\beta}_{1A}\right) \partial_{\tau'} \Lambda^{\ell\alpha,
m\beta}_{cc}  \nonumber \\ [2mm] 
&&\hspace{1.5cm} + \left\{ \Lambda^{\ell\alpha, m\beta}_{cc}
\left(\d^{\ell\alpha}_{jx'} - \d^{m\beta\ast}_{jy'}\right)
\d^{m\beta}_{iy} \Delta^{\ell\alpha}_{ji} \Delta^{m\beta}_{\tilde c
\tilde c} + c.c. \right\}\bigg] .
\eeq
The corresponding proper time expression is given in (A8) of the
Appendix. Note that `c.c.' stands for complex conjugate, as usual.
\item[{(ii)}] Diagrams with a matrix fermion propagator give the
contribution 
\beq
\Gamma_\theta &=& - g \sum \cdots \int \cdots \bigg[ -tr_{\rm dirac} 
\left(\Delta^{\ell\alpha}_{\theta\theta}
\Lambda^{\ell\alpha, m\beta}_{\theta\theta}\right) \Delta^{m\beta}_{AA}
\nonumber \\ [2mm]  
&& + i \ tr_{\rm dirac} \left(\left[\Delta^{\ell\alpha}_{\theta\theta},
\gamma^1\right] \Lambda^{\ell\alpha, m\beta}_{\theta\theta}\right)
\Delta^{m\beta}_{1A} + tr_{\rm dirac} \left(\gamma^k
\Delta^{\ell\alpha}_{\theta\theta} \gamma^i \Lambda^{\ell\alpha,
m\beta}_{\theta\theta}\right) \Delta^{m\beta}_{ik}\bigg] 
\eeq
In the above, we have used an obvious short-hand notation for the
various sums and integrals which are identical to those in (4.3). 
We will use this notation in the following
also. The Dirac trace in the last term is rather tedious to evaluate,
but finally one gets a moderately simple expression for $\Gamma_\theta$  
in the proper time representation. This is given in the Appendix in (A9).
\item[{(iii)}] Diagrams with a matrix gauge propagator give the
contribution
\[
\Gamma_A = \Gamma_{1A} + \Gamma_{2A} + \Gamma_{3A} ,
\]
where 
\beq
\Gamma_{1A} &=& - g \sum \cdots \int \cdots \bigg[ -\partial_{\tau'}
\Delta^{\ell\alpha}_{ji} \partial_\tau \Delta^{m\beta}_{ij} +
\Delta^{\ell\alpha}_{ji} \partial_\tau \partial_{\tau'}
\Delta^{m\beta}_{ij} \nonumber \\ [2mm]
&&\hspace{1cm} - 2 iv\tau \partial_{\tau'} \Delta^{\ell\alpha}_{AA}
\Delta^{m\beta}_{1A} + 2iv\tau \Delta^{\ell\alpha}_{AA}
\partial_{\tau'} \Delta^{m\beta}_{1A} \nonumber \\ [2mm]
&&\hspace{1cm} + v^2 \tau\tau' \Delta^{\ell\alpha}_{1A}
\Delta^{m\beta}_{1A} + \partial_\tau \Delta^{\ell\alpha}_{cc}
\partial_{\tau'} \Delta^{m\beta}_{cc}\bigg] \Lambda^{\ell\alpha,
m\beta}_{AA} , \\
\Gamma_{2A} &=& - g \sum \cdots \int \cdots 
\Lambda^{\ell\alpha, m\beta}_{AA} \left(2\d^{\ell\alpha\ast}_{ix} -
\d^{m\beta}_{iy}\right) \left(2\d^{\ell\alpha}_{jx'} -
\d^{m\beta\ast}_{jy'}\right) \Delta^{\ell\alpha}_{AA}
\Delta^{m\beta}_{ij} , \\
\Gamma_{3A} &=& {1\over 2} g \sum \cdots \int \cdots \Lambda^{\ell\alpha,
m\beta}_{AA} \ tr_{\rm dirac} \left(\Delta^{\ell\alpha}_{\theta\theta} 
\Delta^{m\beta}_{\theta\theta} \right) .
\eeq
These contributions to the \tl effective action have been evaluated   
and are listed in (A10)--(A12) in the Appendix.
\item[{(iv)}] Finally, diagrams with a matrix bosonic propagator give
the contribution
\[
\Gamma_X = \sum^8_{q=1} \Gamma_{qX} ,
\]
where
\beq
\Gamma_{1X} &=& - {1\over 2} g \sum \cdots \int \cdots \ tr_{\rm dirac} 
\left(\Delta^{\ell\alpha}_{\theta\theta} \gamma^i
\Delta^{m\beta}_{\theta\theta} \gamma^j \right) \Lambda^{\ell\alpha,
m\beta}_{ij} , \\
\Gamma_{2X} &=& {1\over 2} g \sum \cdots \int \cdots
\bigg\{\Lambda^{\ell\alpha, m\beta}_{ij} \d^{\ell\alpha}_{jx'}
\d^{m\beta}_{iy} \Delta^{\ell\alpha}_{\tilde c \tilde c}
\Delta^{m\beta}_{\tilde c \tilde c} + c.c. \bigg\} , \\
\Gamma_{3X} &=& - {1\over 2} g \sum \cdots \int \cdots
\Lambda^{\ell\alpha, m\beta}_{ij} \left(\d^{\ell\alpha\ast}_{ix} +
\d^{m\beta}_{iy}\right) \left(\d^{m\beta\ast}_{jy'} +
\d^{\ell\alpha}_{jx'}\right) \times \nonumber \\ [2mm]
&&\hspace{1cm}  \left(\Delta^{\ell\alpha}_{kf} \Delta^{m\beta}_{fk} +
\Delta^{\ell\alpha}_{AA} \Delta^{m\beta}_{AA} - 2
\Delta^{\ell\alpha}_{1A} \Delta^{m\beta}_{1A}\right) , \\
\Gamma_{4X} &=& - g \sum \cdots \int \cdots
\Lambda^{\ell\alpha, m\beta}_{ij} \left(2 \d^{\ell\alpha}_{kx'} - 
\d^{m\beta\ast}_{ky'}\right) \left(2 \d^{\ell\alpha\ast}_{fx} -
\d^{m\beta}_{fy}\right) \Delta^{\ell\alpha}_{ji} \Delta^{m\beta}_{fk}, \\ 
\Gamma_{5X} &=& - {1\over 2} g \sum \cdots \int \cdots \bigg\{
\Lambda^{\ell\alpha, m\beta}_{ij} \left(2 \d^{\ell\alpha}_{kx'} - 
\d^{m\beta\ast}_{ky'}\right) \left(2 \d^{m\beta}_{fy} -
\d^{\ell\alpha\ast}_{fx}\right) \nonumber \\ [2mm]
&&\hspace{1.5cm} \Delta^{\ell\alpha}_{jf} \Delta^{m\beta}_{ik} +
c.c. \bigg\} , \\  
\Gamma_{6X} &=& g \sum \cdots \int \cdots \bigg\{
\Lambda^{\ell\alpha, m\beta}_{ij} \left(\d^{\ell\alpha}_{jx'} + 
\d^{m\beta\ast}_{jy'}\right) \left(2 \d^{\ell\alpha\ast}_{fx} -
\d^{m\beta}_{fy}\right) \nonumber \\ [2mm]
&&\hspace{1.5cm} \Delta^{\ell\alpha}_{ki} \Delta^{m\beta}_{fk} +
c.c. \bigg\} , \\
\Gamma_{7X} &=& -g \sum \cdots \int \cdots \Lambda^{\ell\alpha,
m\beta}_{ij} \bigg\{-4 iv\tau \Delta^{\ell\alpha}_{1A}
\partial_{\tau'} \Delta^{m\beta}_{ij} - 2 iv\tau \partial_{\tau'}
\Delta^{\ell\alpha}_{1A} \Delta^{m\beta}_{ij} \nonumber \\ [2mm]
&&~~~~ + 4 \Delta^{\ell\alpha}_{AA} \partial_\tau \partial_{\tau'}
\Delta^{m\beta}_{ij} + 4 \partial_\tau \Delta^{\ell\alpha}_{AA}
\partial_{\tau'} \Delta^{m\beta}_{ij} + \partial_\tau \partial_{\tau'}
\Delta^{\ell\alpha}_{AA} \Delta^{m\beta}_{ij} \bigg\}, \\
\Gamma_{8X} &=& -g \sum \cdots \int \cdots \Lambda^{\ell\alpha,
m\beta}_{11} \bigg\{10 iv\tau \Delta^{\ell\alpha}_{AA}
\partial_{\tau'} \Delta^{m\beta}_{1A} + 8 iv\tau \partial_{\tau'}
\Delta^{\ell\alpha}_{AA} \Delta^{m\beta}_{1A} \nonumber \\ [2mm]
&&~~~~ - 5\partial_{\tau'} \Delta^{\ell\alpha}_{1A} \partial_\tau 
\Delta^{m\beta}_{1A} + 4 v^2 \tau\tau' \Delta^{\ell\alpha}_{1A}
\Delta^{m\beta}_{1A} - 4 \partial_\tau \partial_{\tau'}
\Delta^{\ell\alpha}_{1A} \Delta^{m\beta}_{1A} \bigg\} .
\eeq
Expressions for these contributions in the proper time representation
are given in (A13)--(A20) in the Appendix.
\end{enumerate}

The total \tl effective action, $\Gamma_{\rm \tl}$, is given by the sum
of the contributions in (4.1)--(4.15).

\section{Ultraviolet Divergences in the \tl Action}

At low velocities and for small values of the parameter $c$, on
general dimensional grounds we expect the \tl effective action to be a
double series of the following form:
\beq
\Gamma_{\rm \tl} &=& {\xi \over c^6} \left( a_1 {b^{11} \over v} + a_2
b^7 v + a_3 b^3 v^3 + d_1 {v^5 \over b} + \cdots \right) \nonumber \\ [2mm]
&& + {\xi \over c^4} \left( a_4 {b^7 \over v} + a_5 b^3 v + d_2 {v^3
\over b} + \cdots \right) \nonumber \\ [2mm]
&& + {\xi \over c^2} \left( a_6 {b^3 \over v} + d_3 {v \over b} + \cdots
\right) \nonumber \\ [2mm]
&& + d_4 {\xi \over bv} + \cdots 
\eeq
Here the quantities $a_1, \cdots , a_6$ and $d_1, \cdots , d_4$ are
pure numbers and $\xi$ is defined in the Appendix in (A1). The
structure of (5.1) follows from the observation that, as defined in
(2.1), the parameter $g$ has dimensions of [length]$^{-3}$. Moreover,
$b^2 , v$ and $c$ all have the dimensions of [length]$^{-2}$ and the
leading term in the \tl calculation goes as $1/vc^6$.

The terms in (5.1) with coefficients $a_1, \cdots , a_6$ all have
positive powers of $b$ with them. On physical grounds all these
coefficients should vanish since they would otherwise give rise to a
potential that diverges at large distances. The terms with
coefficients $d_1 , \cdots , d_4$ all give rise to a potential that
goes as $\rho^{-2}$, where $\rho^2 = b^2 + v^2\tau^2$. This is
precisely the structure of the leading terms, in the limit $c =
f^{-1} \rightarrow 0$, of the \tl effective potential calculated
using the DBI action, given in (6.11) of \cite{DM}. For agreement with
matrix theory, the coefficients of these terms must match between
these two calculations.

We have calculated the various terms in (5.1) using the results
presented in the Appendix.\footnote{Extensive use of Mathematica was
made in these calculations.} We find that while indeed $a_1 = 0$, all
the other coefficients have uncanceled divergences. For example, for
the term proportional to $\xi v/c^6$, we get 
\beq
&& {3\over 4} \bigg[ \int^\infty_0 ds_1 \int^\infty_0 ds_2
~e^{-b^2(s_1+s_2)} \left\{ {(s_1 + s_2)^{3/2}\over s_1^{7/2} s_2^{7/2}}
- {16\over 5} {(s_1 + s_2)^{-1/2} \over s_1^{5/2} s_2^{5/2}} \right\}
\nonumber \\ [2mm]
&& \hspace{1cm} - 2 \int^\infty_0 ds_1 \ {e^{-b^2s_1} \over s^2_1} 
\int^\infty_0 {ds_3 \over  s_3^{7/2}} \bigg] .
\eeq
Similar ill-defined expressions are obtained for all the other
terms. Since the general structure of all the terms is similar to
(5.2), let us see in some detail where the two type of terms in (5.2)
originate.

Consider the second term in (5.2) first. The structure of this term
suggests that it originates from $\Gamma_Z$, (A6). There is actually
another contribution to this term which comes from those terms in the
sum of (4.3)--(4.15) which are proportional to $b^2$. To see this, let
us write the total effective action at 2-loops as
\beq
\Gamma_{\rm \tl} &\equiv& \xi \int^\infty_0 ds_1 \int^\infty_0 ds_2
\int^\infty_0 ds_3 \ e^{-b^2 (s_1 + s_2)} W (s_1, s_2, s_3) \nonumber \\[2mm]
&& + \xi \int^\infty_0 ds_1 \int^\infty_0 ds_2 \ e^{-b^2 (s_1 + s_2)} X
(s_1, s_2) \nonumber \\ [2mm]
&& + \xi \int^\infty_0 ds_1 \int^\infty_0 ds_3 \ e^{-b^2 s_1} Y (s_1, s_3),
\eeq
$W , X$ and $Y$ can be read-off from the results given in the
Appendix. Let us now explicitly separate out the $b^2$ piece in $W$:
\be
W (s_1, s_2, s_3) = \omega (s_1, s_2, s_3) + b^2 {\cal U} (s_1, s_2, s_3) .
\ee
The $b^2$ terms come only from diagrams like Fig. 1 containing a
matrix gauge or bosonic propagator. So the quantity 
${\cal U} (s_1, s_2, s_3)$ can be read-off from the sum of (4.5)--(4.15). 
An explicit expression for it is given in the Appendix in (A21). Now, 
using the identity
\[
b^2 e^{-b^2 (s_1 + s_2)} = - {1\over 2} (\partial_{s_1} +
\partial_{s_2}) e^{-b^2 (s_1 + s_2)} ,
\]
we may rewrite this piece as follows:
\beq
&& \int^\infty_0 ds_1 \int^\infty_0 ds_2 \int^\infty_0 ds_3 \ b^2 e^{-b^2 (s_1 + s_2)} {\cal U} (s_1, s_2, s_3) \nonumber \\[2mm]
&=& {1\over 2} \int^\infty_0 ds_1 \int^\infty_0 ds_2 \int^\infty_0
ds_3 \ e^{-b^2 (s_1 + s_2)} (\partial_{s_1} + \partial_{s_2}) {\cal U} (s_1,
s_2, s_3) \nonumber \\ [2mm]
&& + {1\over 2} \int^\infty_0 ds_1 \int^\infty_0 ds_3 ~e^{-b^2 s_1}
\left[ {\cal U} (s_1, 0, s_3) + {\cal U} (0, s_1, s_3)\right] .
\eeq
Thus, we have 
\beq
\Gamma_{\rm \tl} &=& \xi \int^\infty_0 ds_1 \int^\infty_0 ds_2
\int^\infty_0 ds_3 ~e^{-b^2 (s_1 + s_2)} \tilde W(s_1, s_2, s_3)\nonumber\\[2mm]
&& + \xi \int^\infty_0 ds_1 \int^\infty_0 ds_2 ~e^{-b^2 (s_1 + s_2)} X
(s_1, s_2) \nonumber \\ [2mm]
&& + \xi \int^\infty_0 ds_1 \int^\infty_0 ds_3 ~e^{-b^2 s_1} \tilde Y
(s_1, s_2), 
\eeq
where
\be
\tilde W (s_1, s_2, s_3) \equiv \omega (s_1, s_2, s_3) + {1\over 2}
(\partial_{s_1} + \partial_{s_2}) {\cal U} (s_1, s_2, s_3) ,
\ee
and
\be
\tilde Y (s_1, s_3) \equiv Y (s_1, s_3) + {1\over 2}
\left[ {\cal U} (s_1, 0, s_3) + {\cal U} (0, s_1, s_3)\right] .
\ee
Using (A6) and (A21) we find that 
\beq
\tilde Y (s_1, s_3) &=& {1\over 16} \left( c^3 s_3^{7/2} \sinh^3cs_1 \sinh vs_1
\right)^{-1} \bigg(12 + 13 \cosh 2vs_1 
+ 39 \cosh 2cs_1 \nonumber \\ [2mm]
&& \hspace{.5cm} - 64 \cosh vs_1 \cosh^3cs_1\bigg) 
+ {3\over 4} \left(c s_3^{3/2} \sinh^3cs_1 \sinh vs_1 \sinh^22cs_3 
\right)^{-1} \times \nonumber \\ [2mm]
&& \bigg[\left( 6 + 8 \cosh 2vs_1 + 18\cosh 2cs_1 - 32
\cosh vs_1 \cosh^3cs_1\right) (1 + \cosh 4cs_3) \nonumber \\ [2mm]
&& \hspace{.5cm} + 2 (5-8\cosh vs_1 \cosh cs_1) \sinh 2cs_1 
\sinh 4cs_3 - 3\cosh 2vs_1 + 3\cosh 2cs_1 \bigg]. \nonumber \\ [2mm] 
\eeq
Expanding in powers of $1/c$, to leading order we get
\be
\tilde Y (s_1, s_3) = (4c^6s^3_1 \ s^{7/2}_3 \ \sinh vs_1)^{-1} (51 + 13
\cosh 2vs_1 - 64 \cosh vs_1) + 0 (1/c^4) .
\ee
Now, further expanding this in powers of $v$, we see that the order
$v^{-1}$ term vanishes, but all the higher powers of $v$ have
nonvanishing coefficients. In particular, the order $v/c^6$ term
exactly reproduces the second term in (5.2).

The trick employed in (5.5) has a diagramatic representation. In terms
of diagrams like Fig. 1, the second term in (5.5) results from
shrinking one of the ``column type'' propagators to a point (large
$b^2$ or small $s$ expansion). Doing this to Fig. 1 produces diagrams
of the type shown in Fig. 2, and hence there is mixing of these two
types of contributions, as we have seen above.

The divergence in the second term in (5.2) comes from small values of
the proper time variable, and this suggest that they are ultraviolet
in nature. This can be seen more directly as follows. We have 
seen that this divergence has basically to do with diagrams of
type Fig. 2. A typical contribution of such a diagram is
\be
\int d\tau \int d^3 \vec x \int d^3 \vec y \int d^3 \vec z \ \Delta
(\vec z , \tau ; \vec x , \tau) \Lambda (\vec x, \vec y, \tau ; \vec z,
\vec y, \tau) .
\ee
In writing (5.11) we have ignored the various indices since they are
not relevant to the present discussion. Now, introducing the momentum
space representation for the matrix propagator, 
\[
\Lambda (\vec x, \vec y, \tau; \vec x', \vec y', \tau') = \int {d^3\vec k
\over (2\pi)^3} \int {d\omega \over 2\pi} \ e^{-i\vec k \left({\vec x +
\vec y \over 2} - {\vec x' + \vec y' \over 2}\right) - i \omega (\tau
- \tau')} \tilde\Lambda (\omega, \vec k, \vec x - \vec y) 
\delta^{(3)} \left((\vec x - \vec y) - (\vec x' - \vec y')\right) ,
\]
where
\[
\tilde\Lambda (\omega, \vec k, \vec x)  \equiv \left[ \omega^2
+ \vec k^2 + \vec x^2 \right]^{-1} ,
\]
which can be obtained from (3.15), we may rewrite (5.11) as
\beq
&& \int {d^3\vec k \over (2\pi)^3} \int {d\omega \over 2\pi} \int d^3
\vec x \int d^3 \vec y \int d\tau \ \tilde\Lambda (\omega, \vec k, \vec
x - \vec y) \Delta (\vec x, \tau ; \vec x, \tau) \nonumber \\ [2mm]
&& = \left[ \int {d^3\vec k \over (2\pi)^3} \int {d\omega \over 2\pi}
\int d^3 \vec y \ \tilde\Lambda (\omega, \vec k, \vec y) \right]  
\left[ \int d\tau \int d^3 \vec x \ \Delta (\vec x, \tau ; \vec x,
\tau)\right] . \nonumber 
\eeq
The first factor looks like a massless bosonic loop in 7-dimensional
space-time, which is ultraviolet divergent. The divergence in the
second factor can also be seen to be coming from the high frequency
oscillators in the energy representation of this integral. 

In a similar fashion one can see that the divergences in the first
term in (5.2) are ultraviolet in nature. This term receives
contributions from diagrams of the type shown in Fig. 1 as well as
those in Fig. 3. The general structure of the contribution of the
former type of diagrams, after integrating over the proper time
variable $s_3$ associated with the matrix propagators, is exactly like
those of the latter type of diagrams. In fact, integrating over $s_3$
effectively shrinks the matrix propagator in Fig. 1, thus
producing the diagram in Fig. 3. A typical contribution of these
latter type of diagrams is 
\[
\int d\tau \left[ \int d^3 \vec x \Delta (\vec x, \tau ; \vec x,
\tau)\right] \left[ \int d^3 \vec y \Delta (\vec y, \tau; \vec y,
\tau)\right] .
\]
It is clear from this expression that high frequency fluctuations
produce the divergences seen in the first term in (5.2).

We have discussed above in detail the origin and the structure of the
divergences in the coefficient $a_2$. Similar general discussion
applies to the coefficients $a_3, \cdots, a_6$ and one finds that
these coefficients are also ultraviolet divergent. Somewhat more work
is required for the coefficients $d_1, \cdots, d_4$. This is because
in a naive double expansion in powers of $1/c$ and $v$, the integrals
over the proper time variable $s_3$ in the different contributions to
the \tl effective action diverge at the large $s_3$ end. This happens
essentially because the matrix propagators are ``massless'',
i.e. there is no factor of $e^{-b^2 s_3}$ in the integrand. The
correct procedure is to first integrate over $s_3$ and then perform
the double expansion. Once this is done, one finds that the
coefficients $d_1, \cdots, d_4$ are also ultraviolet divergent.

\section{Concluding Remarks}

In this paper we have presented the results of a calculation of the
\tl effective action for a \dz in the background of the recently
discussed D0-D6 bound state configuration. As we have seen, the \tl
effective action has uncanceled ultraviolet divergences. Similar
ultraviolet divergences are potentially present in the calculation of
the 1-loop effective action also. However, in that calculation, done
in the proper time representation like the present calculation, the
divergent parts precisely cancel in the sum of contributions from
different sectors (boson, fermion, gauge and ghost), leaving behind a
finite answer for the 1-loop effective action. In general, one expects
matrix theory loop calculations to be divergent for the present
background because of nonrenormalizability of 7-dimensional gauge
theory. The ultraviolet divergent \tl result that we have got is
consistent with this expectation. Since a full string theory
calculation is expected to be ultraviolet finite, this means that
massive open string states do not decouple from the 6-brane, as
expected from general considerations \cite{SEN, SEI}.

One of our motivations for the present calculation was the surprising
supergravity/\sym agreement at \tl found in \cite{JB}. In view of our
result and the above considerations, it now seems even more surprising
that a simple \tl effective \sym action should reproduce the
supergravity result. Presumably the effective \sym action can be
obtained from a full string theory calculation. It would be nice to
have a better understanding of this.

\vskip .5in 
\noindent
{\bf Acknowledgements}
\vskip .2in
\noindent
It is a pleasure to thank G. Mandal for numerous discussions and for a
collaboration at an early stage of this work.

\newpage

\centerline{\bf APPENDIX}

\bigskip

In this appendix we list the contributions of various diagrams in the
proper time representation. For this purpose it is useful to define  
$$
{g Q^2_6 \over 64 \sqrt \pi} \equiv \xi ,
\eqno (A1)
$$
$$
{\sinh vs_1 \ \sinh vs_2 \over v \sinh  v(s_1 + s_2)} \equiv h_v , \ \ (s_3 +
h_v)^{-1} \equiv k_v ,
\eqno (A2)
$$
$$
{\sinh cs_1 \ \sinh cs_2 \over c \sinh c (s_1 + s_2)} \equiv h_c , \ \ (s_3 +
h_c)^{-1} \equiv k_c , \left({\tanh2cs_3 \over 2c} + h_c \right)^{-1}
\equiv \tilde k_c ,
\eqno (A3)
$$
$$
{\sqrt{k_v} \over \sinh v (s_1 + s_2)} \ {k^3_c \over c^3 \sinh^3 c (s_1 +
s_2)} \equiv \Omega (s_1, s_2, s_3) ,
\eqno (A4)
$$
$$
{\sqrt{k_v} \over \sinh v (s_1 + s_2)} \ {k_c \tilde k^2_c \over c^3 \sinh^3 c (s_1 +
s_2) \cosh^2 2cs_3} \equiv \tilde\Omega (s_1, s_2, s_3) .
\eqno (A5)
$$
We now list the various contributions.
\beq
\Gamma_Z &=& {9\over 8} \xi \int^\infty_0 ds_1 \int^\infty_0 {ds_3
\over s_3^{7/2}} \ {e^{-s_1 b^2} \over c^3 \sinh^3 cs_1 \sinh vs_1}
(2\cosh^2vs_1 + 3\cosh 2cs_1) \nonumber \\ [2mm]
&& + {3\over 4} \xi \int^\infty_0 ds_1 \int^\infty_0 {ds_3 \over
s_3^{3/2} \sinh^2 2cs_3} \ {e^{-s_1b^2} \over c\ \sinh^3 cs_1 \sinh vs_1} \bigg[
4\cosh^2 vs_1 (1 + 8 \cosh^2 2cs_3) \nonumber \\ [2mm]
&& + 2\cosh 2cs_1 (7 + 20\cosh^2 2cs_3) + 12 \sinh 2cs_1 \sinh 4cs_3 \bigg]
. \hspace{4cm} (A6) \nonumber 
\eeq
We have used $4n=Q_6$ in arriving at this expression. Also, here and in the 
following the first term comes from the $\ell = m$ case and the second term 
from $\ell \neq m$. 
\beq
\Gamma_\phi &=& {1\over 16} \xi \int^\infty_0 ds_1 \int^\infty_0 ds_2
\ e^{-b^2(s_1 + s_2)} (v/\sinh vs_1 \sinh vs_2 \sinh v (s_1 + s_2))^
{1\over 2} (\sinh cs_1 \sinh cs_2)^{-3}  \nonumber \\ [2mm]
&& \times \bigg[ {1\over 2} + 2\cosh 2vs_1 + 6\cosh 2cs_1 
+ {1\over 2} \cosh 2vs_1 \cosh 2vs_2 + {3\over 2} \sinh 2vs_1 \sinh 2vs_2  
\nonumber \\ [2mm] 
&& + 6\cosh 2cs_1 \cosh 2vs_2 + {15\over 2} \cosh 2cs_1 \cosh 2cs_2  
+ {9\over 2} \sinh 2cs_1 \sinh 2cs_2 + s_1 \leftrightarrow s_2 \bigg] 
\nonumber \\ [2mm]
&& + {3\over 16} \xi \int^\infty_0 ds_1 \int^\infty_0 ds_2
\ e^{-b^2(s_1 + s_2)} (v/\sinh vs_1 \sinh vs_2 \sinh v (s_1 + s_2))^
{1\over 2} (\sinh cs_1 \sinh cs_2)^{-3}  \nonumber \\ [2mm]
&& \times \bigg[ {1\over 2} + 2\cosh 2vs_1 + 6\cosh 2cs_1 
+ {1\over 2} \cosh 2vs_1 \cosh 2vs_2 + {3\over 2} \sinh 2vs_1 \sinh 2vs_2 
\nonumber \\ [2mm]
&& + 6\cosh 2cs_1 \cosh 2vs_2 + {15\over 2} \cosh 2cs_1 \cosh 2cs_2 
- {3\over 2} \sinh 2cs_1 \sinh 2cs_2 
+ s_1 \leftrightarrow s_2 \bigg].(A7) \nonumber 
\eeq
In the above, as well as in the following, `$s_1 \leftrightarrow s_2$'
stands for the entire preceeding expression rewritten with this
substitution. 
\beq
\Gamma_C &=& {1\over 32} \xi \int^\infty_0 ds_1 \int^\infty_0 ds_2
\int^\infty_0 ds_3 \ e^{-b^2(s_1 + s_2)} \Omega (s_1, s_2, s_3) \bigg[
vh_v k_v (\sinh 2vs_1 + \coth vs_1 \cosh 2vs_2) \nonumber \\ [2mm]
&& + 6c h_ck_c {\cosh c(s_1 + 2s_2)\over \sinh cs_1} + s_1 \leftrightarrow s_2 \bigg] \nonumber \\ [2mm] 
&&+ {3\over 32} \xi \int^\infty_0 ds_1 \int^\infty_0 ds_2
\int^\infty_0 ds_3 \ e^{-b^2(s_1 + s_2)} \tilde\Omega (s_1, s_2, s_3) \bigg[
vh_v k_v (\sinh 2vs_1 + \coth vs_1 \cosh 2vs_2)  \nonumber \\ [2mm]
&& + 2ch_ck_c {\cosh c(s_1 + 2s_2)\over \sinh cs_1} + 4c h_c \tilde k_c \bigg( \cosh c
(s_1 - 2s_2) \nonumber \\ [2mm]
&& - \sinh c (s_1 - 2s_2) \tanh2cs_3\bigg)/\sinh cs_1 
+ s_1 \leftrightarrow s_2 \bigg] . \hspace{6.1cm} (A8) \nonumber 
\eeq
\beq
\Gamma_\theta &=& - \xi \int^\infty_0 ds_1 \int^\infty_0 ds_2
\int^\infty_0 ds_3 \ e^{-b^2(s_1+s_2)} \Omega (s_1, s_2, s_3) \bigg[
3ch_ck_c \coth cs_1 \cosh cs_1 (\cosh vs_1  \nonumber \\ [2mm]
&& + \cosh v (s_1 + 2s_2)) + {1\over 2} h_vk_v {v\ \cosh^2 cs_1 \over 
\sinh vs_1} (\cosh cs_1 + 3\cosh c (s_1 + 2s_2)) \nonumber \\ [2mm] 
&& + 6ch_ck_c  \coth cs_1 \cosh c (s_1 + 2s_2) \cosh vs_1 + s_1 \leftrightarrow
s_2 \bigg] \nonumber \\ [2mm]
&& - 3 \xi \int^\infty_0 ds_1 \int^\infty_0 ds_2
\int^\infty_0 ds_3 \ e^{-b^2(s_1+s_2)} \tilde\Omega (s_1, s_2, s_3) \bigg[
\bigg(c h_c k_c {\cosh^2c (s_1 + 2s_3) \over \sinh cs_1} \nonumber \\ [2mm]
&& + 2 c h_c \tilde k_c {\coth cs_1 \cosh c (s_1 + 2s_3) \over \cosh 2cs_3}\bigg)
\left( \cosh vs_1 + \cosh v (s_1 + 2s_2)\right) \nonumber \\ [2mm]
&& + {1\over 2} h_vk_v {v \over \sinh vs_1} \bigg(\cosh cs_1 \cosh^2 c(s_1 +
2s_3) + \cosh c (s_1 + 2s_2) \cosh^2 c (s_1 + 2s_3) \nonumber \\ [2mm]
&& + 2\cosh cs_1 \cosh c (s_1 + 2s_3) \cosh c (s_1 - 2s_2 - 2s_3)\bigg) 
\nonumber \\ [2mm]
&& + 2\cosh vs_1 \bigg\{ h_ck_c {c\over \sinh cs_1} \cosh c (s_1 + 2s_3) 
\cosh c(s_1 - 2s_2 -2s_3) + h_c \tilde k_c {c \over \sinh cs_1 \cosh 2cs_3} 
\nonumber \\ [2mm]
&& \times \left( \cosh c (s_1+2s_2) \cosh c(s_1 + 2s_3) 
+ \cosh cs_1 \cosh c (s_1 - 2s_2 - 2s_3)\right) \bigg\} + s_1
\leftrightarrow s_2 \bigg] . \nonumber \\ [2mm]
&& \hspace{15cm} (A9) \nonumber  
\eeq
\newpage
\beq
\Gamma_{1A} &=& - {1\over 16} \xi \int^\infty_0 ds_1 \int^\infty_0 ds_2
\int^\infty_0 ds_3 \ e^{-b^2(s_1+s_2)} \Omega (s_1, s_2, s_3) \bigg[
{v \over \sinh v (s_1+s_2)} \times \nonumber \\ [2mm]
&& (2 + \cosh 2vs_1 \cosh 2vs_2 + 6\cosh 2c(s_1+s_2)) \left( \cosh v (s_1 - s_2) + {k_v \sinh^2 v (s_1 - s_2) \over 4v\ \sinh \ v(s_1+s_2)}\right) 
\nonumber \\ [2mm] 
&& - {v\ \cosh v (s_1-s_2) \over 2\sinh v(s_1+s_2)} (1 + 
\sinh 2vs_1 \sinh 2vs_2) 
+ {1\over 2} k_v\ \sinh 2vs_1 \sinh 2vs_2  \nonumber \\ [2mm] 
&& \times \left( {1\over 4\sinh^2v(s_1+s_2)} -
v^2h_v^2\right) - v (1 - {1\over 2} h_v k_v) {\sinh 2v (s_1-s_2) \sinh v
(s_1-s_2) \over \sinh v (s_1+s_2)}\bigg] \nonumber \\ [2mm]
&& - {3\over 16} \xi \int^\infty_0 ds_1 \int^\infty_0 ds_2
\int^\infty_0 ds_3 \ e^{-b^2(s_1+s_2)} \tilde\Omega (s_1, s_2, s_3) \bigg[
{v \over \sinh v (s_1+s_2)} \times \nonumber \\ [2mm]
&& (2 + \cosh 2vs_1 \cosh 2vs_2 + 2\cosh 2c(s_1+s_2) \nonumber \\ [2mm] 
&& + 4\cosh 2c (s_1-s_2)) \left(\cosh v (s_1 - s_2) + {k_v
\sinh^2 v (s_1 - s_2) \over 4v\ \sinh \ v(s_1+s_2)}\right) \nonumber \\ [2mm]
&& - {v\ \cosh v (s_1-s_2) \over 2\sinh v(s_1+s_2)} (1 + 
\sinh 2vs_1 \sinh 2vs_2) 
+ {1\over 2} k_v\ \sinh 2vs_1 \sinh 2vs_2 \times \nonumber \\ [2mm] 
&& \left( {1\over 4\sinh^2v(s_1+s_2)}- v^2h_v^2\right)
- v (1 - {1\over 2} h_v k_v) {\sinh 2v (s_1-s_2) \sinh v
(s_1-s_2) \over \sinh v (s_1+s_2)}\bigg] . \hspace{1cm} (A10) \nonumber 
\eeq
\beq
\Gamma_{2A} &=& - {1\over 16} \xi \int^\infty_0 ds_1 \int^\infty_0 ds_2
\int^\infty_0 ds_3 \ e^{-b^2(s_1+s_2)} \Omega (s_1, s_2, s_3) \bigg[
{b^2 \over 2} \cosh 2vs_1  \nonumber \\ [2mm]
&& + {v\over 4} \cosh 2vs_1 \cosh 2vs_2 \left({\cosh v (s_1 - s_2)\over \sinh v
(s_1+s_2)} + vh^2_vk_v\right) + {3\over 2} k_c \cosh 2vs_1 \bigg\{{5
\over 2} \cosh 2cs_2 \nonumber \\ [2mm]
&& + cs_3 \coth c(s_1+s_2) + 3ch_c \sinh 2cs_2 - {3\over 2} {\sinh c(s_1-s_2)
\over \sinh c (s_1+s_2)} \cosh 2cs_2 \bigg\} + s_1 \leftrightarrow s_2 \bigg] 
\nonumber \\ [2mm]
&& - {3\over 16} \xi \int^\infty_0 ds_1 \int^\infty_0 ds_2
\int^\infty_0 ds_3 \ e^{-b^2(s_1+s_2)} \tilde\Omega (s_1, s_2, s_3) \bigg[
{b^2 \over 2} \cosh 2vs_1  \nonumber \\ [2mm]
&& + {v\over 4} \cosh 2vs_1 \cosh 2vs_2 \left({\cosh v (s_1 - s_2)\over \sinh v
(s_1+s_2)} + vh^2_vk_v\right) + \cosh 2vs_1 \bigg\{{5
\over 4} k_c \cosh 2cs_2 \nonumber \\ [2mm]
&& + \tilde k_c \cosh 2cs_2 \left({5\over 2} + {1\over 2} \tanh2cs_3 {\cosh c
(s_1-s_2) \over \sinh c (s_1 + s_2)} - 4ch_c \ \tanh2cs_3 \right)\nonumber \\ [2mm]
&& + {1\over 2} cs_3 k_c \coth c(s_1 + s_2) + (2\tanh2cs_3 - 5ch_c) \tilde
k_c \sinh 2cs_2 + {3\over 2} ch_ck_c \sinh 2cs_2 \nonumber \\ [2mm]
&&- {3\over 4} {\sinh c (s_1-s_2) \over \sinh c (s_1+s_2)} \left((k_c +
2\tilde k_c) \cosh 2cs_2 + 2 \tilde k_c \tanh2cs_3 \sinh 2cs_2\right)\bigg\} +
s_1 \leftrightarrow s_2 \bigg] . \hspace{.5cm} (A11) \nonumber 
\eeq
\newpage
\beq
\Gamma_{3A} &=& - {1\over 2} \xi \int^\infty_0 ds_1 \int^\infty_0 ds_2
\int^\infty_0 ds_3 \ e^{-b^2(s_1+s_2)} \Omega (s_1, s_2, s_3) \bigg[
b^2 \cosh v (s_1-s_2) \cosh^3c (s_1+s_2) \nonumber \\ [2mm]
&& + {vh_vk_v \over 2\sinh v (s_1+s_2)} \cosh^3c (s_1+s_2) + 3cs_3k_c
\cosh v(s_1-s_2) \cosh c (s_1+s_2) \coth c(s_1+s_2)\bigg] \nonumber \\ [2mm]
&& - {3\over 2} \xi \int^\infty_0 ds_1 \int^\infty_0 ds_2
\int^\infty_0 ds_3 \ e^{-b^2(s_1+s_2)} \tilde\Omega (s_1, s_2, s_3) \bigg[
b^2 \cosh v (s_1-s_2) \cosh c (s_1+s_2) \times \nonumber \\ [2mm]
&& \cosh^2c(s_1-s_2) + {vh_vk_v \over 2\sinh v(s_1+s_2)} \cosh c(s_1+s_2) 
\cosh^2c (s_1-s_2) + cs_3k_c \cosh v (s_1-s_2) \times \nonumber \\ [2mm] 
&& {\cosh^2c (s_1-s_2)\over \sinh c(s_1+s_2)} 
+ \tilde k_c \tanh2cs_3 \cosh v (s_1-s_2) \cosh c(s_1-s_2) \coth c
(s_1+s_2) \bigg] . \hspace{1cm} (A12) \nonumber  
\eeq
\beq
\Gamma_{1X} &=& - {1\over 2} \xi \int^\infty_0 ds_1 \int^\infty_0 ds_2
\int^\infty_0 ds_3 \ e^{-b^2(s_1+s_2)} \Omega (s_1, s_2, s_3) \bigg[
b^2 \bigg\{\cosh v (s_1-s_2) \cosh c (s_1+s_2) \nonumber \\ [2mm]
&& + 6 \cosh v (s_1+s_2) \cosh c(s_1-s_2)\bigg\} \cosh^2c (s_1+s_2) + {2v \over
\sinh v(s_1+s_2)} \bigg\{ (\cosh c (s_1+s_2) \nonumber \\ [2mm]
&& + 3 \cosh c (s_1-s_2)) \cosh^2c(s_1+s_2) - {3 \over 4} h_vk_v 
\left(\cosh c(s_1+s_2) + 2\cosh c(s_1-s_2)\right) \times \nonumber \\ [2mm]
&& \cosh^2 c(s_1 + s_2) \bigg\} + 3cs_3k_c \coth c(s_1+s_2) 
\bigg\{ 4\cosh v (s_1+s_2) \cosh c (s_1-s_2) \nonumber \\ [2mm]
&& + \left(\cosh v (s_1-s_2) + 2 \cosh v(s_1+s_2)\right) \cosh c(s_1+s_2) 
\bigg\} \bigg] \nonumber \\ [2mm]
&& - {3\over 2} \xi \int^\infty_0 ds_1 \int^\infty_0 ds_2
\int^\infty_0 ds_3 \ e^{-b^2(s_1+s_2)} \tilde\Omega (s_1, s_2, s_3)
\bigg[ b^2 \bigg\{\cosh v (s_1-s_2) \cosh c (s_1+s_2) \times \nonumber \\ [2mm]
&& \cosh^2c(s_1-s_2) + 2\cosh v (s_1+s_2) \cosh c (s_1-s_2) \bigg( \cosh^2c
(s_1-s_2) \nonumber \\ [2mm]
&& + 2\cosh c (s_1+s_2) \cosh c (s_1+s_2+4s_3)\bigg)\bigg\} \nonumber \\ [2mm]
&& + {v \over \sinh v (s_1+s_2)}
\bigg\{ 4 \bigg(\cosh cs_1 \cosh cs_2 \cosh^2 c(s_1-s_2) \nonumber \\ [2mm]
&& + \cosh c (s_1+s_2) \cosh c (s_1-s_2) \cosh c(s_1+s_2+4s_3)\bigg) 
- h_v k_v \bigg( {3\over 2} \cosh c (s_1+s_2) \times \nonumber \\ [2mm]
&& \cosh^2c (s_1-s_2) + \cosh^3c (s_1-s_2) + 2\cosh c (s_1+s_2) \cosh c (s_1-s_2) \cosh c(s_1+s_2+4s_3)\bigg) \bigg\} \nonumber \\ [2mm]
&& + {cs_3k_c \over \sinh c (s_1+s_2)} \bigg\{ \left(\cosh v (s_1-s_2) + 
2\cosh v(s_1+s_2)\right) \cosh^2c (s_1-s_2) \nonumber \\ [2mm]
&& + 4 \cosh v (s_1+s_2) \cosh c (s_1-s_2) \cosh c (s_1+s_2+4s_3)\bigg\} \nonumber \\ [2mm]
&& + {\tilde k_c \tanh2cs_3 \over \sinh c (s_1+s_2)} \bigg\{\left( \cosh v
(s_1-s_2) + 2\cosh v (s_1+s_2)\right) \cosh c (s_1+s_2) \cosh c (s_1-s_2)
\nonumber \\ [2mm]
&& + 2 \cosh v (s_1+s_2) \left(\cosh c (s_1+s_2) \cosh c (s_1+s_2+4s_3) + 
\cosh^2c(s_1-s_2)\right) \bigg\} \bigg] . \hspace{.8cm} (A13) \nonumber
\eeq
\beq
\Gamma_{2X} &=& {1\over 16} \xi \int^\infty_0 ds_1 \int^\infty_0 ds_2
\int^\infty_0 ds_3 \ e^{-b^2(s_1+s_2)} \Omega (s_1, s_2, s_3) \bigg[
b^2 + {v\over 2} \left({\cosh v(s_1-s_2)\over \sinh v (s_1+s_2)} + vh^2_vk_v
\right)  \nonumber \\ [2mm]
&& \hspace{3cm} + 3cs_3 k_c {\cosh c (s_1-s_2) \over \sinh c (s_1+s_2)}\bigg]
\nonumber \\ [2mm]
&& + {3\over 16} \xi  \int^\infty_0 ds_1 \int^\infty_0 ds_2
\int^\infty_0 ds_3 \ e^{-b^2(s_1+s_2)} \tilde\Omega (s_1, s_2, s_3) \bigg[
b^2 + {v\over 2} \left({\cosh v(s_1-s_2)\over \sinh v (s_1+s_2)} +
vh^2_vk_v\right) \nonumber \\ [2mm]
&& \hspace{3cm} + cs_3 k_c {\cosh c (s_1-s_2) \over \sinh c (s_1+s_2)} + 
\tilde k_c \tanh2cs_3 {\cosh c (s_1+s_2+4s_3) \over \sinh c (s_1+s_2)}\bigg]
. \hspace{1cm} (A14) \nonumber 
\eeq
\beq
\Gamma_{3X} &=& - {1\over 8} \xi \int^\infty_0 ds_1 \int^\infty_0 ds_2
\int^\infty_0 ds_3 \ e^{-b^2(s_1+s_2)} \Omega (s_1, s_2, s_3) \bigg[2b^2
+ v \left({\cosh v(s_1-s_2)\over \sinh v (s_1+s_2)} + vh^2_vk_v\right)\nonumber\\[2mm] 
&& + {3\over 2} k_c + 6cs_3k_c {\cosh c (s_1-s_2)\over \sinh c(s_1+s_2)} \bigg]
\left(1+\cosh 2v (s_1+s_2) + 3\cosh 2c (s_1+s_2)\right) \nonumber \\ [2mm]
&& + {3\over 8} \xi  \int^\infty_0 ds_1 \int^\infty_0 ds_2
\int^\infty_0 ds_3 \ e^{-b^2(s_1+s_2)} \tilde\Omega (s_1, s_2, s_3)
\bigg[ 2b^2 + v \left({\cosh v(s_1-s_2)\over \sinh v (s_1+s_2)} +
vh^2_vk_v\right) \nonumber \\ [2mm]
&& + {1\over 2} k_c + 2cs_3k_c {\cosh c (s_1-s_2)\over \sinh c (s_1+s_2)} + \tilde
k_c \cosh 4cs_3 + 2c\ \sinh 4cs_3 \nonumber \\ [2mm]
&& + 2\tilde k_c \tanh2cs_3 \ \cosh 4cs_3 {\cosh cs_1 \cosh cs_2\over \sinh c (s_1+s_2)}
\bigg] \nonumber \\ 
&& \times (1 + \cosh 2v (s_1+s_2) + \cosh 2c (s_1+s_2) + 2\cosh 2c (s_1-s_2))
\hspace{3.5cm} (A15) \nonumber 
\eeq 
\beq
\Gamma_{4X} &=& - {1\over 32} \xi \int^\infty_0 ds_1 \int^\infty_0 ds_2
\int^\infty_0 ds_3 \ e^{-b^2(s_1+s_2)} \Omega (s_1, s_2, s_3) \bigg[(2 +
\cosh 2vs_1 + 6\cosh 2cs_1) \nonumber \\ 
&& \times \left( b^2 + {v\over 2} \cosh 2vs_2  \left({\cosh v(s_1-s_2)\over
\sinh v (s_1+s_2)} + vh^2_vk_v\right)\right) + 3cs_3k_c \coth c(s_1+s_2)
\nonumber \\  
&& + {3\over 2} k_c \cosh 2cs_2 \left(5 - 3 {\sinh c (s_1-s_2) \over
\sinh c(s_1+s_2)} \right) + 9ch_ck_c \ \sinh 2cs_2 + s_1 \leftrightarrow s_2 \bigg]
\nonumber \\ [2mm] 
&& - {3\over 32} \xi  \int^\infty_0 ds_1 \int^\infty_0 ds_2
\int^\infty_0 ds_3 \ e^{-b^2(s_1+s_2)} \tilde\Omega (s_1, s_2, s_3)
\bigg[(2 + \cosh 2vs_1 + 2\cosh 2cs_1  \nonumber \\ [2mm] 
&& + 4\cosh 2c (s_1 + 2s_3))) \left( b^2 + {v\over 2} \cosh 2vs_2  \left({\cosh v(s_1-s_2)\over \sinh v (s_1+s_2)} + vh^2_vk_v\right)\right) 
\nonumber \\ [2mm] 
&& + cs_3k_c \ \coth c(s_1+s_2) + {1\over 2} k_c \cosh 2cs_2 
\left(5 - 3 {\sinh c (s_1-s_2) \over \sinh c(s_1+s_2)} \right) + 
3ch_ck_c \ \sinh 2cs_2 \nonumber \\ [2mm] 
&& + \tilde k_c \cosh 2cs_2 \tanh2cs_3 \bigg({\cosh c(s_1-s_2) \over \sinh c (s_1+s_2)} - 8ch_c \bigg) + \tilde k_c \tanh2cs_3 \sinh 2cs_2 \times
\nonumber \\ [2mm]
&& \left(4 -3 {\sinh c (s_1-s_2) \over \sinh c (s_1+s_2)}\right) 
+ \tilde k_c \cosh 2cs_2 \left(5-3 {\sinh c(s_1-s_2) \over 
\sinh c(s_1+s_2)}\right) - 10 ch_c \tilde k_c \sinh 2cs_2 \nonumber \\ [2mm] 
&& + s_1 \leftrightarrow s_2 \bigg] . \hspace{12.3cm} (A16) \nonumber
\eeq
\beq
\Gamma_{5X} &=& - {1\over 16} \xi \int^\infty_0 ds_1 \int^\infty_0
ds_2 \int^\infty_0 ds_3 \ e^{-b^2(s_1+s_2)} \Omega (s_1, s_2, s_3)\bigg[
b^2 + {v\over 2} \left({\cosh v(s_1-s_2)\over \sinh v (s_1+s_2)} +
vh^2_vk_v\right) \nonumber \\ [2mm]
&& \times \cosh 2vs_1 \ \cosh 2vs_2 + 3cs_3k_c {\cosh c (s_1-s_2)\over \sinh c (s_1+s_2)} -
6k_c \cosh 2c (s_1-s_2) \bigg] \nonumber \\ [2mm]
&& - {3\over 16} \xi  \int^\infty_0 ds_1 \int^\infty_0 ds_2
\int^\infty_0 ds_3 \ e^{-b^2(s_1+s_2)} \tilde\Omega (s_1, s_2, s_3)
\bigg[b^2 + {v\over 2} \left({\cosh v(s_1-s_2)\over \sinh v (s_1+s_2)} +
vh^2_vk_v\right) \nonumber \\ [2mm]
&& \times \cosh 2vs_1 \ \cosh 2vs_2 + cs_3k_c {\cosh c (s_1-s_2)\over \sinh c (s_1+s_2)} -
2k_c \cosh 2c (s_1-s_2) \nonumber \\ [2mm]
&& + \tilde k_c \tanh2cs_3 \left({\cosh c(s_1-s_2) \over \sinh c (s_1+s_2)} + 
10ch_c\right) \cosh 2c (s_1+s_2+2s_3) \nonumber \\ [2mm]
&& - 4 \tilde k_c \cosh 2c (s_1+s_2+2s_3) - 2 c (5-9 h_c \tilde k_c) \sinh 2c
(s_1+s_2+2s_3) \bigg] . \hspace{2.3cm} (A17) \nonumber
\eeq
\beq
\Gamma_{6X} &=& {1\over 8} \xi \int^\infty_0 ds_1 \int^\infty_0 ds_2
\int^\infty_0 ds_3 \ e^{-b^2(s_1+s_2)} \Omega (s_1, s_2, s_3) \bigg[2b^2
+ v\ \cosh 2vs_1 \ \cosh 2vs_2 \bigg(vh^2_vk_v \nonumber \\ [2mm]
&& + {\cosh v(s_1-s_2) \over \sinh v (s_1+s_2)}\bigg) 
+ 6cs_3k_c {\cosh c (s_1-s_2) \cosh 2c(s_1+s_2)\over \sinh c(s_1+s_2)} \nonumber \\ [2mm]
&& + {3\over 2} k_c \ \cosh 2c (s_1+s_2) + 9ch_c\ k_c\ \sinh 2c(s_1+s_2)\bigg]
\nonumber\\[2mm]  
&& + {3\over 8} \xi  \int^\infty_0 ds_1 \int^\infty_0 ds_2
\int^\infty_0 ds_3 \ e^{-b^2(s_1+s_2)} \tilde\Omega (s_1, s_2, s_3)
\bigg[ 2b^2 + v\ \cosh 2vs_1 \ \cosh 2vs_2 \bigg(vh^2_vk_v \nonumber \\ [2mm]
&& + {\cosh v(s_1-s_2) \over \sinh v (s_1+s_2)}\bigg) 
+ 2cs_3k_c {\cosh c (s_1-s_2) \cosh 2c(s_1+s_2)\over \sinh c(s_1+s_2)} \nonumber \\[2mm]
&& + {1\over 2} k_c \cosh 2c (s_1+s_2) + 3ch_c k_c \sinh 2c (s_1+s_2) + \tilde
k_c \cosh 4cs_3 \cosh 2c(s_1-s_2) \nonumber \\ [2mm]
&& - 3\tilde k_c \sinh 4cs_3 {\sinh c (s_1-s_2) \sinh 2c (s_1-s_2) \over
\sinh c(s_1+s_2)} + 2ch_c\tilde k_c \sinh 4cs_3 \ \cosh 2c(s_1-s_2) \nonumber \\
&& + \tilde k_c \tanh 2cs_3 \bigg( 2\cosh 4cs_3 {\cosh3c (s_1-s_2)\over \sinh c (s_1+s_2)}
+ ch_c {\sinh c (2s_1-s_2+4s_3) \over \sinh cs_2} \nonumber \\ [2mm] 
&& + ch_c {\sinh c (2s_2-s_1+4s_3) \over \sinh cs_1}\bigg)\bigg] 
\hspace{9.2cm} (A18) \nonumber  
\eeq
\beq
\Gamma_{7X} &=& - {1\over 64} \xi \int^\infty_0 ds_1 \int^\infty_0 ds_2
\int^\infty_0 ds_3 \ e^{-b^2(s_1+s_2)} \Omega (s_1, s_2, s_3) \bigg[ 
v (2 + \cosh 2vs_2 + 6\cosh 2vs_2)  \nonumber \\ [2mm]
&& \times \bigg\{ - 2\sinh 2vs_1 {\sinh v (s_1-s_2) \over \sinh v (s_1+s_2)} +
\cosh 2vs_1 {\cosh v(s_1-s_2) \over \sinh v (s_1+s_2)} \nonumber \\ [2mm]
&& + vh^2_vk_v \bigg(2\sinh 2vs_1 (\coth vs_1 + 2\coth vs_2) 
+ \cosh 2vs_1 (\coth^2vs_1 + 4\coth^2vs_2 \nonumber \\ [2mm]
&& + 4\coth vs_1 \coth vs_2)\bigg) \bigg\} + s_1 \leftrightarrow s_2 \bigg]
\nonumber \\ [2mm]
&& - {3\over 64} \xi \int^\infty_0 ds_1 \int^\infty_0 ds_2
\int^\infty_0 ds_3 \ e^{-b^2(s_1+s_2)} \tilde\Omega (s_1, s_2, s_3) \bigg[
v (2 + \cosh 2vs_2 + 2\cosh 2cs_2  \nonumber \\ [2mm]
&& + 4\cosh 2c (s_2+2s_3)) \bigg\{ - 2\sinh 2vs_1 {\sinh v (s_1-s_2) \over 
\sinh v (s_1+s_2)} +
\cosh 2vs_1 {\cosh v(s_1-s_2) \over \sinh v (s_1+s_2)} \nonumber \\ [2mm]
&& + v h^2_v k_v \bigg(2\sinh 2vs_1 (\coth vs_1 + 2\coth vs_2) + 
\cosh 2vs_1 (\coth^2vs_1 + 4\coth^2vs_2 \nonumber \\ [2mm]
&& + 4\coth vs_1 \coth vs_2)\bigg) \bigg\} + s_1 \leftrightarrow s_2 \bigg] . 
\hspace{8cm} (A19) \nonumber 
\eeq
\beq
\Gamma_{8X} &=& - {1\over 64} \xi \int^\infty_0 ds_1 \int^\infty_0
ds_2 \int^\infty_0 ds_3 \ e^{-b^2(s_1+s_2)} \left[\Omega (s_1, s_2, s_3)
+ 3 \tilde\Omega (s_1, s_2, s_3)\right] \nonumber \\ [2mm]
&& \times v \bigg[ 2\cosh 2vs_1 \sinh 2vs_2 {\sinh v (s_1-s_2) \over \sinh v (s_1+s_2)}
- 5 \sinh 2vs_1 \sinh 2vs_2 {\cosh v(s_1-s_2) \over \sinh v (s_1+s_2)} \nonumber \\ [2mm]
&& - vh^2_vk_v \bigg\{ 2\cosh 2vs_1 \sinh 2vs_2 (4\coth vs_1 + 5\coth vs_2) -
\sinh 2vs_1 \sinh 2vs_2 (4\coth^2vs_1 \nonumber \\ [2mm]
&& + 5\coth vs_1 \coth vs_2 -4) \bigg\} + s_1 \leftrightarrow s_2 \bigg] .
\hspace{7.5cm} (A20) \nonumber
\eeq
\beq
{\cal U} (s_1, s_2, s_3) &=& - {1\over 16} \Omega (s_1, s_2, s_3)
\bigg[ 1 + \cosh 2vs_1 + 2\cosh 2v (s_1+s_2) + 3\cosh 2cs_1 \nonumber \\ [2mm]
&& + 6\cosh 2c (s_1+s_2) + 8\cosh v (s_1-s_2) \cosh^3c(s_1+s_2) 
\nonumber \\ [2mm]
&& + 24\cosh v (s_1+s_2) \cosh c (s_1-s_2) \cosh^3c (s_1+s_2)
+ s_1 \leftrightarrow s_2 \bigg] \nonumber \\ [2mm]
&& - {3\over 16} \tilde\Omega (s_1, s_2, s_3) \bigg[ 1 + \cosh 2vs_1 +
2\cosh 2v (s_1+s_2) + \cosh 2cs_1 \nonumber \\ [2mm]
&& + 2\cosh 2c (s_1+2s_3) + 2\cosh 2c (s_1+s_2) + 4\cosh 2c (s_1-s_2) 
\nonumber \\ [2mm]
&& + 8\cosh v (s_1-s_2) \cosh c (s_1+s_2) \cosh^2c (s_1-s_2) 
\nonumber \\ [2mm]
&&+ 16\cosh v (s_1+s_2) \cosh c (s_1+s_2+4s_3) \cosh c (s_1+s_2) 
\cosh c(s_1-s_2) \nonumber \\ [2mm] 
&& + 8 \cosh v (s_1+s_2) \cosh^3c (s_1-s_2) +  s_1 \leftrightarrow s_2 \bigg] .
\hspace{4.7cm}  (A21) \nonumber 
\eeq


\newpage
\begin{thebibliography}{999}
\bibitem{SHEIN} H.J. Sheinblatt, {\it Phys. Rev.} {\bf D57} (1998) 2421, 
hep-th/9705054.
\bibitem{WATI} W. Taylor, {\it Nucl. Phys.} {\bf B508} (1997) 122, 
hep-th/9705116.
\bibitem{KVK} E. Keski-Vakkuri and P. Kraus, 
{\it Nucl. Phys.} {\bf B510} (1998) 199, hep-th/9706196.
\bibitem{JMP} J.M. Pierre, {\it Phys. Rev.} {\bf D56} (1997) 6710,
hep-th/9707102.
\bibitem{BISY} A. Brandhuber, N. Itzhaki, J. Sonnenschein and
S. Yankielowicz, {\it Phys. Lett.} {\bf B423} (1998) 238, hep-th/9711010.
\bibitem{DM} A. Dhar and G. Mandal, {\it Nucl. Phys.} {\bf B531} (1998),
hep-th/9803004.
\bibitem{NI} N. Itzhaki,  hep-th/9809063.
\bibitem{BFSS} T. Banks, W. Fischler, S. Shenker and L. Susskind,
Phys. Rev. {\bf D55} (1997) 5112, hep-th/9610043.
\bibitem{GK} G. Gibbons and R. Kallosh, {\it Phys. Rev.} {\bf D51}
(1995) 2839.
\bibitem{LP} H. Lu and C. N. Pope, hep-th/9606047.
\bibitem{DoM} P. Dobiasch and D. Maison, {\it Gen. Relativ.
Gravitation} {\bf 14} (1982) 231.
\bibitem{CD} A. Chodos and S. Detweiler, {\it Gen. Relativ.
Gravitation} {\bf 14} (1982) 879.
\bibitem{DP} D. Pollard, {\it J. Phys.} {\bf A16} (1983) 565.
\bibitem{GW} G. Gibbons and D.L. Wiltshire, {\it Ann. Phys.} {\bf
167} (1986) 201.
\bibitem{GL} G. Lifschytz, {\it Nucl. Phys.} {\bf B520} (1998) 105, 
hep-th/9612223.
\bibitem{CT} I. Chepelev and A.A. Tseyltin, 
{\it Nucl. Phys.} {\bf B515} (1998) 73, hep-th/9709087.
\bibitem{JB} J. Branco, {\it Class. Quant. Grav.} {\bf 15} (1998) 3739,
hep-th/9806186.
\bibitem{SEN} A. Sen, {\it Adv. Theor. Math. Phys.} {\bf 2} (1998) 51,
 hep-th/9709220.
\bibitem{SEI} N. Seiberg, {\it Phys. Rev. Lett.} {\bf 79} (1997) 3577, 
hep-th/9710009.
\end{thebibliography}
\end{document}